# Contact identification for assembly disassembly simulation with a haptic device

Robert Iacob, Peter Mitrouchev, Jean-Claude Léon

*G-SCOP Laboratory, INP Grenoble-UJF-CNRS,*

*46, av. Félix Viallet, 38031, GRENOBLE Cedex 1, FRANCE*

Phone: +33 4 76 82 51 27      Fax: +33 4 76 57 46 95

Robert.Iacob@gmail.com, Peter.Mitrouchev@g-scop.inpg.fr, Jean-Claude.Leon@inpg.fr

**Abstract**:

Assembly/disassembly (A/D) simulations using haptic devices are facing difficulties while simulating insertion/extraction such as cylinders from holes. In order to address this configuration as well as others, an approach showing that contact identification between components can efficiently contribute either to a new A/D simulation preparation process relying on two types of shape representations (mesh and CAD NURBS models), or directly to the real time simulation process when it is performed with 6D haptic devices, is presented in this paper. The model processing pipeline is described and illustrated to show how information can be propagated and used for contact detection. Then, the contact identification process is introduced and illustrated through an example.

## 1. Introduction

Assembly/Disassembly (A/D) simulation of industrial products finds a strong interest in interactive simulations through immersive and real-time schemes. The relative mobility of components is a key element contributing to A/D simulations based on 3D component models. This mobility can be represented either exactly with translations, rotations and helices or approximated with infinitesimal translations only. They can be also deduced from the relative positions of the components or strictly specified by the user.

Many scientific contributions focus on A/D simulations in Virtual Reality (VR), addressing the product or some of its subsystems rather than an isolated component (Graf, 2002). Raghavan et al., (1999) described an interactive tool for evaluating assembly sequences using augmented reality. These immersive environments are rapidly developing and the representation of the components used there is of polyhedral type. Recently, Zhong et al., (2005), presented a constraint-based methodology for intuitive and precise solid modeling in a VR environment. A constraint reasoning engine is also developed to automatically deduce allowable motions. Haptics is not used there and the model of mobility is purely for design rather than A/D purposes.

Sun et al., (2002) proposed an interactive task planner that incorporates a two-handed VR interface for assembly of CAD models but issues related to haptics are not addressed. In Jung, (2003) the author proposed a knowledge-based model of connection-sensitive part features and algorithms that define the task-level interface for A/D simulations. There, natural language is the key element of the user interface rather than haptics.

An approach was set up by Liu et al., (2007) that realizes assembly relationship identification, constraint solution and constrained motion guidance for interactive assembly in VR. Models, imported in IGES or SAT format, contain only geometric information, hence a necessary step before virtual assembly is to define the assembly port, which describes connection interface between components. Indeed, the user must interactively specify these connection ports by selecting a collection of geometric entities and providing other relevant parameters. However, haptic devices and their specific needs are not incorporated in the proposed approach. Edwards et al., (2004) developed a system to evaluate collisions detected between various parts. The most significant example is the insertion of a bolt into a hole. Using the same example, Lim et al., (2007) and Howard et al. (2007), performed separate tests to evaluate collision detection and the amount of clearance needed has been calculated. In order to reduce the effect of these approximations and to obtain a proper virtual simulation, a complete contact model is required, which shows the importance of model processing for haptic simulations.

We can mention that a few approaches have been developed for contacts finding using polyhedral models. Because of the local nature of the algorithms, their accuracy and robustness are low and they are sensitive to the triangle sizes (Coma, 2003).

Prior work has focused on a simulation framework (Iacob et al., 2007) capable of addressing interactive as well as real time A/D simulations. However, the specific constraints needed for haptic simulations where not incorporated.

In this context, the aim of this work consists in introducing a subset of a simulation framework contributing to A/D simulations using haptics. The use of 3D component models is closely related to contacts between them. The proposed framework fully automates contact identification, providing a smarter way to monitor collisions and take into account the advantages of haptics in A/D simulations.

The paper is organized as follows. Section 2 gives some concepts about contact identification and highlights the contributions of the proposed method. Then, section 3 presents the model processing pipeline of the current approach and the steps of the assembly model processing through the A/D simulation preparation phase. Based on the model preparation phases, the main treatments contributing to the contact identification are described. The contact identification operators are then described at section 4. Finally, examples of assembly processing and contacts identification are detailed in section 5.

## 2. Contribution of contact identification to VR A/D simulation

Two configurations of contact identifications are considered. The first one takes place during the model preparation stage, as an off-line process, and produces information used during the insertion/extraction phases of components. The second one is performed during the real time manipulation of components when they collide with each other. The proposed method can interact with kinematic models used in haptic arms because the contact type and the nature of the surfaces involved in a contact can help characterizing the nature and the kinematic parameters between two components in contact (Hamri et al., 2006). This is a complement to the geometric location of contacts to express the effective relative movements between neighboring components. We note that this is very helpful for VR simulations addressing the simulation of maintainability operations, when haptic devices are used. There, it is important to avoid side effects due to configurations where surfaces of two distinct components are close to each other, e.g. when a cylinder is inserted/extracted into/from a hole. In such a configuration, collision detection algorithms are based on polyhedral models of the components and many collisions can occur depending on the respective positions of the cylindrical surfaces, thus generating unacceptable vibrations, e.g. during the insertion phase of one component into another when their nominal dimensions are equal. On the other hand, having the correct kinematic mobilities between the components would allow the servo control of the haptic device to generate the trajectories as specified by the corresponding kinematic joint (planar fit, cylindrical fit, …), thus avoiding the undesired effects and improving its usage.

As a result, automating the characterization of the relative mobility of components is a key element contributing to A/D simulations. This relative mobility strongly relies on functional contacts between components, i.e. contact involving surfaces of type plane, cylinder, cone, sphere, as well as information that can be extracted from CAD models and propagated to the meshes needed for VR simulations. Having all the contacts between components identified, the simulation can be performed in better circumstances, thus providing a smarter way to detect collisions. Moreover, using information about surfaces, the real time search for new contacts can be easily performed. This represents a better solution to eliminate vibrations of haptic devices. The following section describes the main stages of the model processing pipeline and its interactions with the contact identification process and the models involved in a VR simulation with haptics.

## 3. Model processing pipeline for A/D simulation with haptics

When the input model of components comes from CAD software, it is important to take advantage of their B-Rep NURBS description to strengthen the algorithms and to obtain a more transparent access to the behaviours of the assembly components during A/D simulations (Iacob et al., 2007). The proposed A/D simulation is particularly suited for industrial applications where a product is composed from an assembly of independent parts. In the model processing workflow (see Fig. 1), the component models are acquired through a STEP file coming from a CAD software.

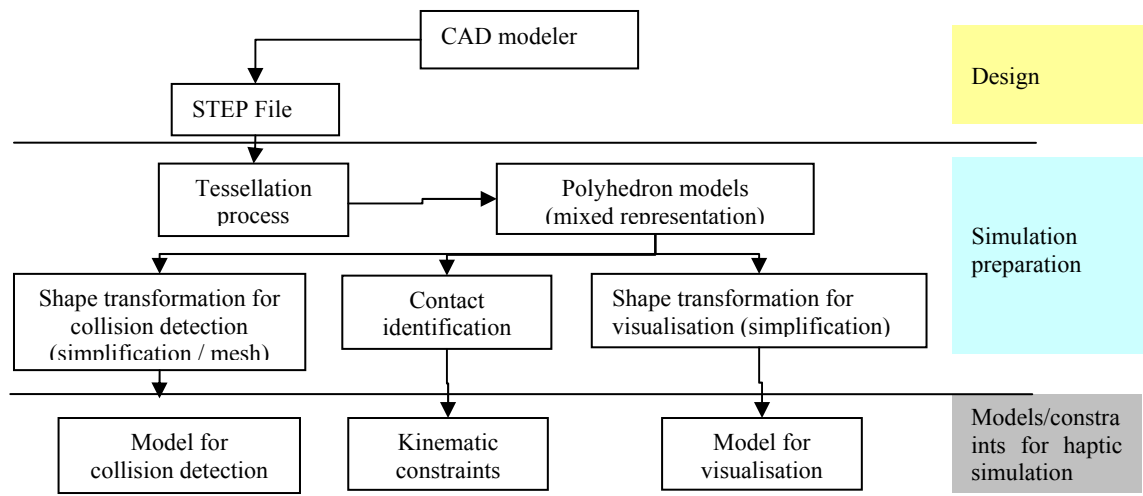

Figure 1. Pipeline of the model processing scheme used to prepare A/D simulation and incorporating contact identification

The proposed approach depends on three data categories considered as mandatory for A/D simulation: the component models, their relative positions and their functional surfaces (planes, cylinders, cones, spheres). The STEP exchange format is robust and efficient for transferring component shapes and it is interesting to notice that these data are only geometric and are provided through the STEP files available with current CAD systems. There, all the information about functional surfaces is available in addition to B-Rep NURBS geometry, the contacts being related to the functional surfaces. Then, the tessellation process taking place allows the user to monitor this process through edge length, chordal deviation and equilaterality constraints. The algorithm set up is free of influence of the tolerances of CAD modeler, i.e. it produces automatically a conform polyhedron whatever the tolerance on the CAD modeler generating the STEP file. Based on the polyhedron generated for each component, NURBS and analytic surfaces parameters are automatically attached to it to form the so-called mixed representation of each component. The mixed representation is the basis for the propagation of the information attached to the B-Rep NURBS models.
Apart from the B-Rep NURBS representation, a facetted representation is necessary in order to visualize the assembly model. In the present case, shape transformations can take place to simplify a component while propagating information about the B-Rep NURBS surfaces every time it is possible. A distance criterion is applied to keep the simplified model within a user-defined envelope. Based on the same initial model, the model suited for collision detection can be generated through shape simplifications (possibly different from the ones performed for the visualization model) and mesh quality constraints. The same distance criterion can be used to generate these collision detection models to keep the consistency with the visualization ones. Shape simplifications can be performed in order to improve the efficiency of the collision detection algorithms while staying compatible with the objectives of the A/D simulation.

Using again the same initial model, the contact identification process takes place and will be the basis for producing the kinematic constraints that will be used to monitor the haptic arm during the insertion/extraction phases. The proposed pipeline can generate these models in an automated manner once accuracy criteria and objectives of shape simplification have been specified and expressed in a procedural manner. A special data structure has been developed, within this framework, which is well suited to link together the different shape representations, i.e. the B-Rep NURBS model, the polyhedron models for visualisation and for collision detection. A subset of this data structure is called High Level Topology (HLT) detailed in (Hamri et al., 2006) and acts as a layer placed on top of the topological description of polyhedral and B-Rep NURBS models both. This HTL combined with the geometry of polyhedrons and B-Rep NURBS models form a mixed shape representation able to take into account and propagate semantic information attached to the input CAD model. Partitions, i.e. a finite set of polyhedron faces matching a B-Rep NURBS patch, are an example of entities contributing to the propagation between polyhedrons and CAD entities.

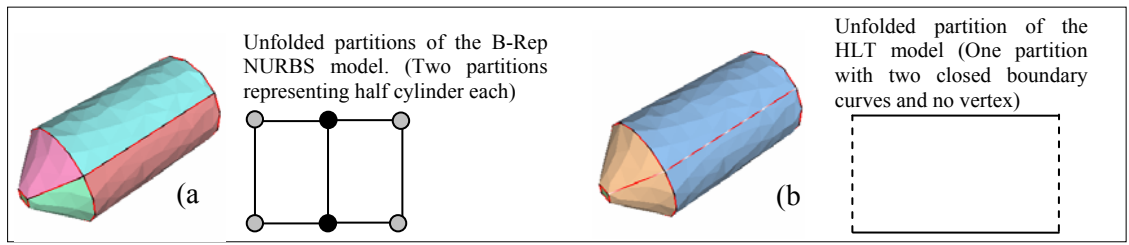

Figure 2. Example of configuration where B-Rep NURBS partitions and edges are merged to describe intrinsically cylindrical surfaces: a) partitions of the B-Rep NURBS model mapped onto the polyhedron model; b) partition obtained after merging

In order to contribute to the contact identification process, CAD modellers are restricted to surface decompositions such that each edge must be adjacent either to two faces if the surface is closed or one face if it is located at the boundary of an open surface. As a consequence, cylindrical surfaces of a shaft are decomposed into two faces whereas the meaning of these areas should be described by only one partition bounded by a closed curve without vertices (see Fig. 2). Indeed, no vertices should be located along a circle because all the points share exactly the same neighbourhood. Other configurations contributing to the description of functional areas and contact descriptions could be listed to show how the limitations of CAD can be overcome. Before starting the contact identification module, the procedure consists in generating the maximal partitions over the boundary of each component. As a result, partitions having the same semantic parameters are merged, e.g. adjacent cylinders having same axis and radius, adjacent planes having the same position and normal, etc. Fig. 2 gives an example of configuration where the HLT data structure can be used to characterize explicitly the partition defining a functional surface. This figure highlights the intrinsic representation of the functional area obtained with the proposed data structure. It is obvious that all the exported/imported CAD models have been created within some tolerance. In order to take into account these inaccuracies, the partition merging module uses two tolerances: a linear and an angular one. At the end of this operator, a List with the Merged Partitions (LMP) is added into the data structure for the contact identification process. This list is then used by the contact identification operator.

Based on the pipeline described at Fig. 1, the front cover of the electric motor (see Fig. 6) is used to illustrate some of the steps of the pipeline prior to the contact identification process. Once the user-specified tolerances have been input, these stages are obtained automatically with the propagation mechanisms based on the HLT data structure and the mixed representation. The different stages of the processing pipeline concerning the front cover of the electric motor (see Fig. 6) are presented in Fig. 3: a) after tessellation with partitions mapped from the CAD model; b) after partition merging; c) partitions highlighting the type of the analytic surfaces (pink: plane, blue: cylinder, orange: cone).

Fig. 4 depicts other steps of the processing pipeline regarding the generation of the polyhedral models needed for the A/D simulation with haptics. Fig. 4a shows the polyhedral model devoted to visualization purposes and generated from the one of Fig. 3b. Fig. 4b illustrates the polyhedral model for collision detection. There, equilaterality and size constraints have been used to produce it. In addition, to exemplify the fact that models can be targeted for specific simulations while staying consistent with each other, different shape simplifications have been performed on the front cover compared to Fig. 4a. They are justified because the through holes removed are smaller than the through holes used by the screws (see Fig. 6), therefore the screws cannot penetrate these holes during the A/D sequence for mounting/dismounting the front cover.

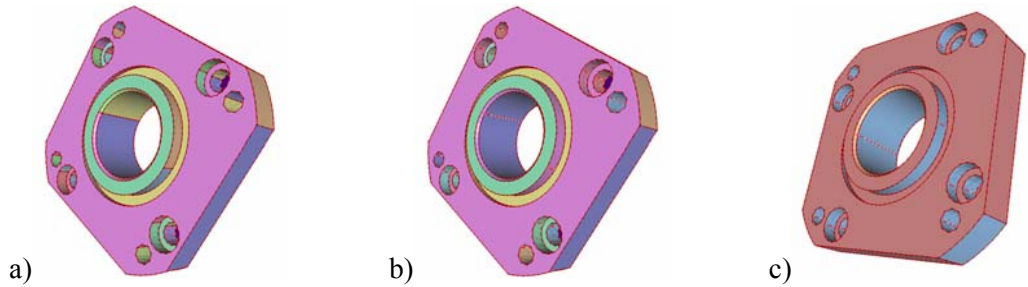

a) b) c)

Figure 3. a) after tessellation; b) after partition merging; c) partitions highlighting the type of analytic surfaces

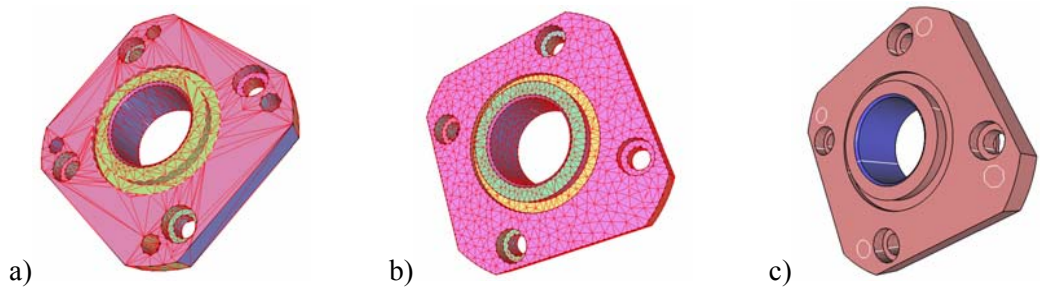

a) b) c)

Figure 4. a) after simplification for visualization; b) after simplification to generate the model for collision detection; c) the collision detection model with partitions tagged for invalidation

Even further, Fig. 4c gives an example where the surfaces involved in the insertion/extraction operations are tagged so that they are not part of the collision detection model once these operations have started. There, the relative movement between the seal or the bearing and the front cover is reduced to translations along and rotations around the axis of cylinder. Indeed, it becomes a kinematic constraint for the haptic device, no longer requiring the collision detection in the area of the tagged surfaces. Taking advantage of this information, it is possible to further speed up the collision detection and simultaneously improve the behavior of the haptic device with a kinematic constraint somehow 'global' rather than purely local ones.

## 4. The contact identification operator

Because A/D simulations in VR use polyhedral representations, few approaches have been developed for contacts finding using these models. The assessment of the state of art research allows us to conclude that the current VR simulation software can be further improved because the results are not always accurate and robust due to the local nature of the algorithms, which are sensitive to triangle sizes, as it was mentioned at section 1. Our simulation framework automates the contact identification and offers a more robust approach to the usage of haptic devices. Thanks of the developed software, at present, the operator can identify five types of contacts: Planar Fit (PLF), Cylindrical Joint (CLJ), Cylindrical Joint unidirectional (CJU), Spherical Fit (SPF) and Linear Annular Fit (LAF), i.e. it is equivalent to the case of a sphere moving inside a cylinder of same diameter.

At present, the user has to select the components in the assembly tree built from the STEP file where the contacts identification should be performed. If two components at least are selected, the application generates a List of Bodies intersecting each other (LBiB). To this end, the bounding box of each component is used in order to check the intersection between bodies and speed up the process. Then, using the LMP created by the partition merging operator (see section 3), four Lists with possible Contacts (LpC) are created for each type of surface. They enumerate:

- plane: coplanar surfaces with collinear and opposite normals,
- cylinder: cylindrical surfaces with same radius and axis and opposite normals,
- cone: conical surfaces with same angle, axis, apex and opposite normals,
- sphere: spherical surfaces with same centre, radius, and opposite normals.

At this point, having all the necessary information structured, the contact identification is performed. For each type of contact, the surfaces from the LpC which belongs to the bodies from the LBiB are tested and a corresponding list is created:

- Plane Fit: planar surfaces with a non empty common area (LcAPP),
- Cylindrical Joint: cylindrical surfaces sharing a common area (LcPVG),
- Cylindrical Joint unidirectional: conical surfaces sharing a common area (LcPGU),
- Spherical Fit: spherical surfaces from LpC and LBiB, knowing that there cannot be more than two spherical surfaces having the same parameters (LcRTL),
- Linear Annular Fit: spherical (convex) surface being inside of a cylindrical (concave) surface (LcLNA).

Partial contacts are also addressed but for the sake of conciseness, they are not described here. Briefly, they can be characterized by configurations where surfaces of revolution are limited to less than 180° in the common contact area. The contacts identification process is fully automated. The list with all the contacts is obtained in only three steps. Any complete assembly previously designed using a CAD software can be virtually checked or analyzed very quickly using the proposed method. For each step some default parameters – adequate for most of the files – are already provided and the process needs only a few seconds to complete.

Similarly, the polyhedral models produced and enriched with the mixed representation are a first step to detect and adapt the behaviour of a haptic device on the fly when collisions are identified. Using the surface types and their relative positions, it is possible to deduce kinematic constraints, thus improving also the haptic device monitoring and avoiding undesired vibrations.

## 5. Results

To illustrate the proposed approach, we have selected the assembly of a DC electric motor, BL17 series type, from Dynetic Systems (see Fig. 6). The electric motor was chosen because it is a common and relevant example and it contains a sufficient number of components: axis – 1 piece; rotor-front – 1 piece; rotor-back – 1 piece; magnets – 4 pieces same dimensions; bearing-front – 1 piece, bearing-back – 1 piece; seal – 1 piece; stator – 1 piece; body – 1 piece; back-cover – 1 piece; front-cover – 1 piece and screws – 4 pieces (same dimensions).

In order to generate the list with all the contacts of the product, three steps must be followed. Figure 5a shows the complete assembly imported in the preparation model framework after the tessellation process. Before starting the contacts identification process, a surface merging operation is required, its result is showed in figure 5b. At this stage all the surfaces, from all the components, having the same parameters are merged to form the partitions of maximal areas.

Using all these data, the list with all the contacts is created. The algorithm presented in the previous section is used and the results are: 21 – Cylindrical joints (CLJ) and 14 – Planar fits (PLF).

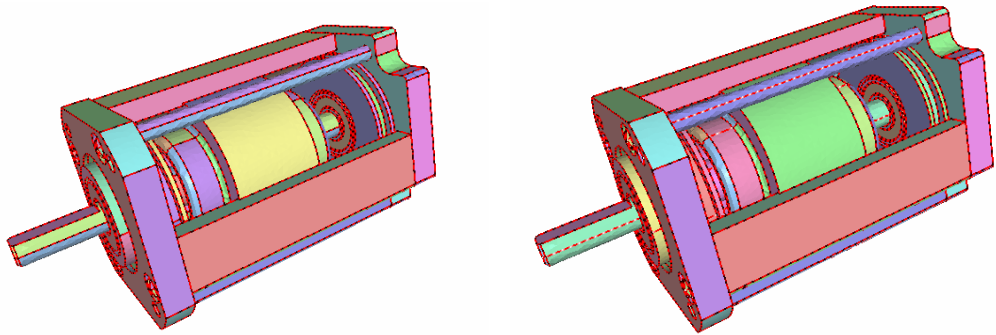

Figure 5.  Electric motor assembly: a) after tessellation, b) after partition merging

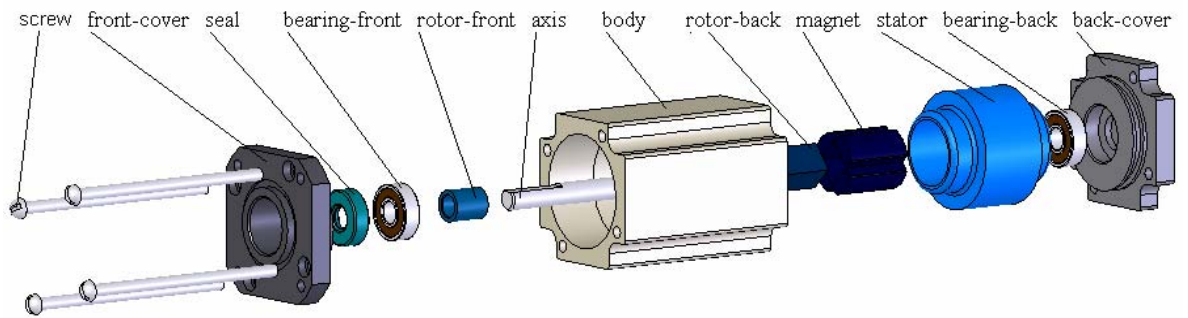

Figure 6.  DC electric motor, BL17 series

The Planar fit contacts are:
Rotor-back      ↔   Magnets (4)      → 4 (PLF ).
Rotor-front     ↔   Rotor-back       → 1 (PLF );   Bearing-front   → 1 (PLF ).
Back-cover      ↔   Bearing-back     → 1 (PLF ).
Body            ↔   Back-cover       → 1 (PLF );   Stator          → 1 (PLF );
                    Front-cover      → 1 (PLF );
Screw (4)       ↔   Front-cover      → 4 (PLF ).

The Cylindrical joint contacts are:
Axis            ↔   Rotor-front      → 1 (CLJ );   Rotor-back      → 1 (CLJ );
                    Bearings (2)     → 2 (CLJ );   Seal            → 1 (CLJ ).
Back-cover      ↔   Bearing-back     → 1 (CLJ ).
Front-cover     ↔   Bearing-front    → 1 (CLJ );   Seal            → 1 (CLJ ).
Stator          ↔   Body             → 1 (CLJ ).
Screw (4)       ↔   Body             → 4 (CLJ );   Back-cover      → 4 (CLJ );
                    Front-cover      → 4 (CLJ ).

## 6. Conclusions

In this paper, a model processing pipeline and a framework for contact identification have been presented. The proposed method provides a smarter way to manage collisions, using the contacts information. The model processing is largely automated and in only three steps the assembly is analyzed and all the information about contacts is stored in a data structure. The contact identification operators combined with the topological description of partitions result in the efficiency of the current approach. The pipeline has been illustrated for each step to show how the polyhedron models and kinematic constraints can be generated and one complete example was presented in order to validate the method.

Further developments of the approach will address a wider range of contacts and the identification of the effective common area between components will be a source of improvement for collision detection and kinematic constraints usage with haptic devices.